\documentclass[manuscript]{aastex6}

\begin{document}

\title{An advanced N-body model for interacting multiple stellar systems}

\author{Miroslav Bro\v z\altaffilmark{1}}
\affil{Astronomical Institute of the Charles University,
Faculty of Mathematics and Physics,\\
V Hole\v sovi\v ck\'ach 2,
CZ-18000 Praha 8,
Czech Republic}

\begin{abstract}
We construct an advanced model for interacting multiple stellar systems
in which we compute all trajectories with a numerical N-body integrator,
namely the Bulirsch--Stoer from the SWIFT package.
We can then derive various observables:
astrometric positions,
radial velocities,
minima timings (TTVs),
eclipse durations,
interferometric visibilities,
closure phases,
synthetic spectra,
spectral-energy distribution,
and even complete light curves.
We use a modified version of the Wilson--Devinney code for the latter,
in which the instantaneous true phase and inclination of the eclipsing binary
are governed by the N-body integration.
If one has all kinds of observations at disposal, a joint $\chi^2$ metric
and an optimisation algorithm (a~simplex or simulated annealing) allows
to search for a global minimum and construct very robust models
of stellar systems. At the same time, our N-body model is free from artefacts
which may arise if mutual gravitational interactions among all components
are not self-consistently accounted for.
Finally, we present a number of examples showing dynamical effects
that can be studied with our code and we discuss how systematic errors
may affect the results (and how to prevent this from happening).
\end{abstract}



\section{Introduction} \label{sec:intro}

Traditional models of eclipsing binaries have to often account
for additional external bodies, most importantly as
a third light, which makes depths of primary and secondary minima shallower;
a light-time effect, causing periodic variations on $O-C$ diagrams;
a precession of the argument of periastron~$\omega$,
shifting the secondary minimum due to perturbations by the 3rd body;
or changes of the inclination~$i$ with respect to the sky plane,
in other words disappearing eclipses.

While analytical theories exist for a description of dynamical perturbations
in triple stellar systems and corresponding transit timing variations
(also known as TTVs, ETVs; see e.g. Brown 1936, Harrington 1968,
S\"oderhjelm 1975, Breiter \& Vokrouhlick\'y 2015, Borkovits et al. 2016),
we would prefer a more general approach --- to account for {\em all\/}
observational data; or at least as many as feasible.
So, our aim is to incorporate
astrometric or speckle-interferometric positions,
radial velocities,
minima timings,
eclipse durations,
spectro-interferometric visibilities,
closure phases,
synthetic spectra,
spectral-energy distribution,
and light curves too.
At the same time, we do not want to be limited by inevitable
approximations of the analytical theories (the N-body problem is not integrable)
and the only way out seems to be an N-body integrator
(as in Carter et al. 2011).

Another aspect is we cannot use analytical photometric models
(like those used for exoplanet transits; Mandel \& Agol 2002, Carter et al. 2008, P\'al 2012),
because the respective simplifications are not acceptable
for stellar eclipses, not speaking about ellipsoidal variations
outside eclipses.

In principle, our approach should be rather straightforward:
we merge two codes into a single one;
namely Levison \& Duncan (1974) SWIFT code,
and Wilson \& Devinney (1971) WD code.
In practice, a lot of work has to be done,
because both of them have to be modified,
we need to extract and derive observable quantities,
read observational data and check them by means of a~$\chi^2$~statistics.
Last but not least, we need to run a minimisation algorithm on top of them.

Even though we do not present new observational data here,
there is one recent application of our N-body model to $\xi$ Tauri quadruple system
which was described in a great amount of detail in Nemravov\'a et al. (2016).
Moreover, there is a comparison with a number of traditional,
observation-specific models. In this `technical' paper,
we prefer to show mostly results of numerical simulations,
or even negative results contradicting the observations,
to demonstrate a sensitivity of our model.

We have a few motivations to do so:
  (i)~no complete and fully self-consistent N-body model exists yet,
      which can account for that many observational constraints,
 (ii)~we improved the model significantly compared to Nemravov\'a et al.
      as we can now fit also complete light curves
      and optionally individual spectra (to be matched by synthetic ones);
(iii)~the previous paper was a bit lengthy and there was simply not enough room
      for a more technical description of our code;
 (iv)~we have to discuss the role of systematics, an experience gained
      during modelling of real multiple stellar systems.


\section{Model description}\label{sec:model}

Let us begin with a description of the numerical integrator and
the photometric model; then we present principal equations,
a definition of the $\chi^2$ metric used to compare the model
with observational data, and a~list of dynamical effects
that can be modelled.

\subsection{Numerical integrator}

\begin{figure}
\centering
\includegraphics[width=8cm]{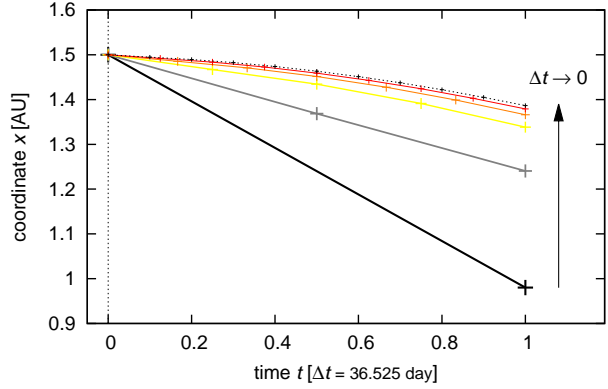}
\caption{A principle of the Bulirsch--Stoer integrator.
There is the time~$t$ as a~independent variable on the abscissa
and one of the coordinates $x_{\rm b}$ on the ordinate.
A series of integrations with decreasing time steps
$\Delta t_i = {\Delta t\over 2}, {\Delta t\over 4}, {\Delta t\over 6}, \dots$ is performed
and then extrapolated for $\Delta t \to 0$ using a rational function.
At the same time, relative differences between successive iterations
have to be smaller than $\epsilon_{\rm BS}$.}
\label{fig:BS}
\end{figure}

We use the Bulirsch--Stoer numerical integrator (Press et al. 1999),
with an adaptive time step, controlled by a unit-less parameter $\epsilon_{\rm BS}$.
The integrator sequentially divides the time step $\Delta t$
by factors 2, 4, 6, \dots, checks if the relative difference
between successive divisions is less than $\epsilon_{\rm BS}$
and then performs an extrapolation $\Delta t\to 0$
by means of a rational function (see Figure~\ref{fig:BS}).
If the maximum number of divisions $n_{\rm max} = 10$ is reached,
the basic time step $\Delta t$ has to be decreased,
with another maximum number of trials $n_{\rm try} = 30$.
We beg to recall this well-known principle here as it is important
to always understand principles and limitations of numerical methods in use.
This kind of integrator is quite general and there are
no restrictions for magnitudes of perturbations, so we can handle
keplerian orbits, tiny N-body perturbations or even violent close encounters.
Even though it is not symplectic, it does not suffer from
an artificial periastron advance. On long time scales,
it is worth to check the energy conservation and eventually decrease
$\epsilon_{\rm BS}$, perhaps down to~$10^{-11}$.

Apart from the internal time step, a user can choose the output
time step $\Delta t_{\rm out}$. The time stepping was adapted
so that we first prepare a list of `times of interest' (corresponding to all observations)
and the integrator outputs coordinates and velocities at exactly these times.
Consequently, the need for additional interpolations is eliminated,
except for minima timings and eclipse durations,
where a linear interpolation from two close neighbouring points
separated by the expected duration is used,
and optionally for light curves (see below).

\subsection{Photometric model}

The only restriction for the geometry of the stellar system,
is that only bodies 1 and 2 may be components of an eclipsing binary
(or an ellipsoidal variable). Nevertheless, there can be any number
of additional bodies, which do contribute to the total light,
but we do not compute eclipses for them.

For light curve computations, we use the WD 2005 version,
in order to produce compatible and comparable results to Phoebe 1.0 (Pr\v sa \& Zwitter, 2005),
but we plan to upgrade in the future.
In brief, the WD code accounts for:
black-body radiation or the Kurucz atmospheres,
bolometric limb darkening,
gravity darkening,
reflection,
an axial rotation, or
the Rossiter--McLaughlin effect.
This is a relatively complex photometric model
(more complex than analytical models of Mandel \& Agol 2002, Carter et al. 2008).
We use {\em no\/} spots or circumstellar clouds in this version.
Usually, the code is called with
mode~0 (no constraints on potentials)
or~2 (the luminosity $L_2$ of the secondary is computed from the temperature $T_2$).
Note a number of parameters in lc.in input file are useless (e.g. orbital elements,
precession and period rates, luminosities, potentials etc.) because they
are driven from elsewhere.

To speed up light curve computations, we can use a binning of times
$\Delta t_{\rm bin}$ and then linearly interpolate light curve points
to the times of observations. For high-cadence data, we can possibly gain
a factor of 10 or 100 speed-up this way, but we have to be sure
there is no physical process in our model which could change magnitudes
on the timescale shorter than~$\Delta t_{\rm bin}$.

\subsection{Principal equations}

Principal equations of our N-body model can be summarized as follows
(the notation is described in~Table~\ref{tab:quantities}) ---
the equation of motion:%
\footnote{The program, including sources and example input data,
is available at \url{http://sirrah.troja.mff.cuni.cz/~mira/xitau/}.}
\begin{equation}
\ddot{\bf r}_{{\rm b}i} = -\sum_{j\ne i}^{N_{\rm bod}} {Gm_j\over r_{ji}^3}\,{\bf r}_{ji} + {\bf f}_{\rm tidal} + {\bf f}_{\rm oblat} + {\bf f}_{\rm ppn}\,,
\end{equation}
with the lowest-order tidal term (Hut 1981):
\begin{equation}
{\bf f}_{\rm tidal} = -\sum_{j\ne i}^{N_{\rm bod}} 3 k_{{\rm L}i} {G m_j^2\over m_i} {R_i^5\over r_{ij}^8} {\bf r}_{ji}\,,\label{eq:tidal}
\end{equation}
oblateness:
\begin{equation}
{\bf f}_{\rm oblat} = -\sum_{j\ne i}^{N_{\rm bod}} {1\over 2} k_{{\rm L}j}\, \omega_{{\rm rot}\,j}^2 {R_i^5\over r_{ij}^5} {\bf r}_{ji}\,,\label{eq:oblat}
\end{equation}
and parametrized post-newtonian (PPN; Mardling \& Lin 2002) terms:
\begin{eqnarray}
{\bf f}_{\rm ppn} &=& -\sum_{j>i}^{N_{\rm bod}} {G(m_j\!+\!m_i)\over r_{ij}^2 c^2} \biggl\{ -2(2-\eta_{ij})\dot r_{ij} \dot{\bf r}_{ij} \nonumber\\
&& +\, \biggl[ (1+3\eta_{ij})\dot{\bf r}_{ij}\!\cdot\dot{\bf r}_{ij} - {3\over 2}\eta_{ij}\dot r_{ij}^{\,2} \nonumber\\
&& -\, 2(2+\eta_{ij}) {G(m_j\!+\!m_i)\over r_{ij}} \biggr] {{\bf r}_{ij}\over r_{ij}} \biggr\} \,.\label{eq:ppn}
\end{eqnarray}
Apart from trivial sky-plane positions~$x_{{\rm b}i}$, $y_{{\rm b}i}$ and radial velocities~$v_{z{\rm b}i}$,
we can derive a number of {\em dependent\/} quantities, such as
mid-eclipse timings (including light-time effects):
\begin{equation}
t_{\rm ecl}' = t_{\rm min} + {z_{{\rm b}1+2} - z_{{\rm b}1+2}(t = T_0)\over c} - {z_{{\rm h}2}\over c} \,,
\end{equation}
eclipse durations:
\begin{equation}
\epsilon'_{\rm ecl} = {2\over\bar v_{{\rm h}2}}\sqrt{(R_1+R_2)^2-\Delta_{\rm min}^2}\,,\label{eq:duration}
\end{equation}
luminosities (alternatively, using a black-body approximation, $\pi B_\lambda(T_{\rm eff})$):
\begin{equation}
L_{j}(T_{{\rm eff}j}, R_j) = 4\pi R_j^2 \int_{\lambda-\Delta\lambda/2}^{\lambda+\Delta\lambda/2} F_{\rm syn}\!\left(\lambda, T_{{\rm eff}\,j}, \log g_j, v_{{\rm rot}\,j}, {\cal Z}_j\right) {\rm d}\lambda\,,\label{eq:blackbody}
\end{equation}
a~limb-darkened complex visibility
(Hanbury Brown et al. 1974;
$\Theta = \pi\theta_j\! \sqrt{u^2\!+\!v^2}$,
$\alpha = 1 - u_{\rm limb}$,
$\beta = u_{\rm limb}$):
\begin{eqnarray}
V'(u, v) &=& \sum_{j=1}^{N_{\rm bod}} {L_j\over L_{\rm tot}} \left({\alpha\over 2} + {\beta\over 3}\right)^{\!\!-1} \Biggl[ \alpha {J_1(\Theta)\over\Theta}+\, \beta\sqrt{\pi\over 2} {J_{3/2}(\Theta)\over\Theta^{3/2}} \Biggr] \, {\rm e}^{-2\pi{\rm i}(u x_{{\rm a}j} + v y_{{\rm a}j})}\,,\label{eq:visibility}
\end{eqnarray}
with $u_{\rm limb}(\lambda, T_{{\rm eff}\,j}, \log g_j, {\cal Z}_j)$ interpolated from Van Hamme 1993);
a~complex triple product:
\begin{equation}
\kern-.3cm T_3' = V'(u_1,v_1) V'(u_2,v_2) V'(-(u_1{+}u_2),-(v_1{+}v_2))\,,\kern-.1cm
\end{equation}
the true phase of the eclipsing binary (at a~time $t$ modified by the light-time effects):
\begin{equation}
\varphi_{\rm ecl}' = {1\over 2\pi}\arctan {\hat O\cdot\hat Y\over \hat O\cdot\hat X}\,,
\end{equation}
its inclination:
\begin{equation}
i_{\rm ecl}' = \arccos({-\hat O\cdot\hat Z})\,,
\end{equation}
Kopal potential (for the WD code which outputs relative magnitudes~$m_V'$):
\begin{equation}
\Omega_{{\rm Kopal}\,j} \doteq \left\langle {1\over r_1} + {q\over r_2} + {1\over 2} (1+q) r_3^2 \right\rangle_{{\rm \!circle\,}{R_j\over r_{12}},\,{\rm where}}\label{eq:kopal}
\end{equation}
$$
{\bf r}_1 \in {\rm circle},\,
{\bf r}_2 = {\bf r}_1 - (1,0,0),\,
{\bf r}_3 = (x_1-{\textstyle{q\over 1+q}},y_1,0),\,
$$
a~normalized synthetic spectrum (with appropriate Doppler shifts):
\begin{equation}
I_\lambda' = \sum_{j=1}^{N_{\rm bod}} {L_j\over L_{\rm tot}} \,I_{\rm syn}\!\left[\lambda\left(1-{v_{z{\rm b}j+\gamma}\over c}\right), T_{{\rm eff}\,j}, \log g_j, v_{{\rm rot}\,j}, {\cal Z}_j\right],
\end{equation}
or a~spectral-energy distribution (in any of the UBVRIJHK bands):
\begin{equation}
F_V' = \sum_{j=1}^{N_{\rm bod}} \left({R_j\over d}\right)^{\!2} \int_0^\infty F_{\rm syn}\!\left[\lambda, T_{{\rm eff}\,j}, \log g_j, v_{{\rm rot}\,j}, {\cal Z}_j\right] f_V(\lambda) {\rm d}\lambda\,,
\end{equation}
$$m_V' = -2.5\log_{10} {F_V'\over F_{V{\rm calib}} \int_0^\infty f_V(\lambda){\rm d}\lambda}\,,$$
where the component spectra (both $I_{\rm syn}$ and $F_{\rm syn}$)
can be either user-supplied, or interpolated on the fly with Pyterpol (Nemravov\'a et al. 2016)
from AMBRE, POLLUX, BSTAR, OSTAR or PHOENIX grids
(Palacios et al. 2010, de Laverny et al. 2012, Lanz \& Huben\'y 2007,
Lanz \& Huben\'y 2003, Husser et al. 2013).

Internally, we use a barycentric left-handed Cartesian coordinate system with
$x$~negative in the right-ascension direction,
$y$~positive in declination, and
$z$~positive in radial, i.e. away from the observer;
the units are day, au, au/day and ${\rm au}^3/{\rm day}^2$
for the time, coordinates, velocities and masses, respectively.
We also need additional coordinate systems, namely:
Jacobian (for computations of hierarchical orbital elements),
1-centric (for an eclipse detection),
1+2~photocentric, or 1+2+3~photocentric (for a comparison with astrometric observations of components~3 and~4).

One may immediately note a minor caveat of our model:
the geometric radius (in Eq.~(\ref{eq:duration})),
the effective radius (in Eq.~(\ref{eq:blackbody})),
the limb-darkened radius (a.k.a.~$\theta_j$ in Eq.~(\ref{eq:visibility})), and
the average radius (used in Eq.~(\ref{eq:kopal}))
are all assumed to be approximately the same.
If this does not hold, it would be necessary to add some three more equations
describing relations between them.

\begin{table}
\caption{Notation used for coordinates, velocities, and a number of other quantities
and uncertainties, which we use in our N-body model.}
\begin{tabular}{ll}
$N_{\rm bod}$                               & number of bodies \\
$m$                                         & mass ($GM_\odot$ units) \\
$q = {m_1\over m_2}$                        & mass ratio \\[-1pt]
$\eta_{ij} = {m_j m_i\over(m_j\!+\!m_i)^2}$ & symmetrized mass ratio \\[-3pt]
$k_{\rm L}$                                 & Love number \\
$\omega_{\rm rot}$                          & rotational angular velocity \\
$x_{\rm b}, y_{\rm b}, z_{\rm b}$           & barycentric coordinates \\
$v_{x{\rm b}}, v_{y{\rm b}}, v_{z{\rm b}}$  & barycentric velocities \\
$x_{\rm h}, y_{\rm h}, z_{\rm h}$           & 1-centric coordinates \\
$v_{x{\rm h}}, v_{y{\rm h}}, v_{z{\rm h}}$  & 1-centric velocities \\
$x_{\rm p}, y_{\rm p}$                      & 1+2 photocentric sky-plane coordinates \\
$x_{\rm p3}, y_{\rm p3}$                    & 1+2+3 photocentric coordinates \\
$x_{\rm a} = {x_{\rm h}\over d}, y_{\rm a}$ & 1-centric coordinates in an angular measure \\
$\hat X, \hat Y, \hat Z$                    & unitvectors aligned with 1+2 eclipsing pair \\
$\hat O = (0,0,-1)$                         & observers direction \\
$\gamma$                                    & systemic velocity \\
$v_{\rm rad}$                               & observed radial velocity \\
$t_{\rm ecl}$                               & mid-epoch of an eclipse of 1+2 pair \\
$\epsilon_{\rm ecl}$                        & eclipse duration \\
$L, L_{\rm tot}$                            & component luminosity and the total one \\
$T_{\rm eff}$                               & effective temperature \\
$R$                                         & stellar radius \\
\end{tabular}
\label{tab:quantities}
\end{table}

\setcounter{table}{0}
\begin{table}
\caption{Cont.}
\begin{tabular}{ll}
$\lambda$, $\Delta\lambda$                  & effective wavelength and bandwidth \\
$B_\lambda(T)$                              & the Planck function \\
$V$                                         & complex visibility, squared visibility is $|V|^2$ \\
$T_3$                                       & complex triple product, closure phase is $\arg T_3$ \\
$u, v$                                      & projected baselines (expressed in cycles, $B\over\lambda$) \\
$\theta = {2R\over d}$                      & angular diameter \\
$u_{\rm limb}$                              & linear limb-darkening coefficient \\
$d$                                         & distance to the system \\
$m_V$                                       & magnitude (in V band or another)\\
$m_0$                                       & zero point \\
$I_\lambda$, $I_{\rm syn}$                  & normalized monochromatic intensity \\
$F_{\rm syn}$                               & absolute monochromatic flux (in ${\rm erg}\,{\rm s}^{-1}\,{\rm cm}^{-2}\,{\rm cm}^{-1}$) \\
$F_{V{\rm calib}}$                          & calibration flux \\
$f_V$                                       & filter transmission coefficient \\
$g = {GM\over R^2}$                         & surface gravity, $\log g$ in cgs \\
$v_{\rm rot}$                               & projected rotational velocity \\
${\cal Z}$                                  & metallicity \\
$\sigma_{\rm sky\,major,\,minor}$           & uncertainty of the astrometric position, \\
                                            & angular sizes of the uncertainty ellipse \\
$\phi_{\rm ellipse}$                        & position angle of the ellipse \\
${\bf R}(\dots)$                            & the corresponding $2\times 2$ rotation matrix \\
$\sigma_{\rm rv}$                           & uncertainty of the radial velocity \\
$\sigma_{\rm ttv}$                          & uncertainty of the eclipse mid-epoch timing \\
\end{tabular}
\end{table}

\setcounter{table}{0}
\begin{table}
\caption{Cont.}
\begin{tabular}{ll}
$\sigma_{\rm ecl}$                          & uncertainty of the eclipse duration \\
$\sigma_{\rm vis}$                          & uncertainty of the squared visibility \\
$\sigma_{\rm clo}$                          & uncertainty of the closure phase \\
$\sigma_{\rm t3}$                           & uncertainty of the triple product amplitude \\
$\sigma_{\rm lc}$                           & uncertainty of the light-curve data \\
$\sigma_{\rm syn}$                          & uncertainty of the normalized intensity \\
$\sigma_{\rm sed}$                          & uncertainty of the spectral-energy distribution \\
$m_j^{\rm min}, m_j^{\rm max}$              & minimum and maximum masses \\
\end{tabular}
\end{table}

\subsection{Observational data}

When we compare our model with observations, we can compute $\chi^2$ for
astrometric positions,
radial velocities,
minima timings (TTVs),
eclipse durations,
interferometric squared visibilities,
closure phases,
triple product amplitudes,
light curves,
synthetic spectra, and
spectral-energy distribution:
\begin{eqnarray}
\chi^2 &=& \underline{\chi^2_{\rm sky}} + \chi^2_{\rm rv} + \chi^2_{\rm ttv} + \chi^2_{\rm ecl} + \underline{\chi^2_{\rm vis}} + \underline{\chi^2_{\rm clo}} + \chi^2_{\rm t3} + \nonumber\\
&& +\, \underline{\chi^2_{\rm lc}} + \underline{\chi^2_{\rm syn}} + \underline{\chi^2_{\rm sed}} + \chi^2_{\rm mass}\,,\label{eq:chi2}
\end{eqnarray}
where:
\begin{equation}
\chi^2_{\rm sky} = \sum_{j=1}^{N_{\rm bod}} \sum_{i=1}^{N_{{\rm sky}\,j}} \left\{ {(\Delta x_{ji})^2 \over \sigma_{{\rm sky\,major}ji}^2} + {(\Delta y_{ji})^2 \over \sigma_{{\rm sky\,minor }ji}^2}\right\}\,, \label{chisky}
\end{equation}
\begin{equation}
(\Delta x_{ji}, \Delta y_{ji}) = {\bf R}\left(-\phi_{\rm ellipse}- {\pi\over2}\right) \times
\pmatrix{
x_{{\rm p}\,ji}' - x_{{\rm p}\,ji} \cr
y_{{\rm p}\,ji}' - y_{{\rm p}\,ji}
}\,,
\end{equation}
\begin{equation}
\chi^2_{\rm rv} = \sum_{j=1}^{N_{\rm bod}} \sum_{i=1}^{N_{{\rm rv}\, j}} {\left(v_{z{\rm b}\,ji}' +\gamma-v_{{\rm rad}\,ji}\right)^2 \over \sigma_{{\rm rv}\,ji}^2}\,,\label{chirv}
\end{equation}
\begin{equation}
\chi^2_{\rm ttv} = \sum_{i=1}^{N_{\rm ttv}} {\left(t'_{{\rm ecl}\,i} - t_{{\rm ecl}\,i}\right)^2 \over \sigma_{{\rm ttv}\,i}^2}\,,
\end{equation}
\begin{equation}
\chi^2_{\rm ecl} = \sum_{i=1}^{N_{\rm ecl}} {\left(\epsilon'_{{\rm ecl}\,i} - \epsilon_{{\rm ecl}\,i}\right)^2 \over \sigma_{{\rm ecl}\,i}^2}\,,
\end{equation}
\begin{equation}
\chi^2_{\rm vis} = \sum_{i=1}^{N_{\rm vis}} {\left({|V'(u_i, v_i)|}^2 - |V|^2_i\right)^2 \over \sigma_{{\rm vis}\,i}^2}\,, \label{chivis}
\end{equation}
\begin{equation}
\chi^2_{\rm clo} = \sum_{i=1}^{N_{\rm clo}} {\left(\arg T_{3i}' - \arg T_{3i}\right)^2 \over \sigma_{{\rm clo}\,i}^2}\,, \label{chiclo}
\end{equation}
\begin{equation}
\chi^2_{\rm t3} = \sum_{i=1}^{N_{\rm t3}} {\left(|T_{3i}|' - |T_{3i}|\right)^2 \over \sigma_{{\rm t3}\,i}^2}\,, \label{chit3}
\end{equation}
\begin{equation}
\chi^2_{\rm lc} = \sum_{k=1}^{N_{\rm band}} \sum_{i=1}^{N_{{\rm lc}\,k}} {\left(m_{Vki}' + m_{0\,k} - m_{Vki}\right)^2 \over \sigma_{{\rm lc}\,ki}^2}\,, \label{chilc}
\end{equation}
\begin{equation}
\chi^2_{\rm syn} = \sum_{i=1}^{N_{\rm syn}} {\left(I_{\lambda\,i}' - I_{\lambda\,i}\right)^2 \over \sigma_{{\rm syn}\,i}^2}\,, \label{chisyn}
\end{equation}
\begin{equation}
\chi^2_{\rm sed} = \sum_{i=1}^{N_{\rm sed}} {\left(m_{Vi}' - m_{Vi}\right)^2 \over \sigma_{{\rm sed}\,i}^2}\,. \label{chised}
\end{equation}
Again, the quantities are described in Table~\ref{tab:quantities}.
The index~$i$ always corresponds to observational data,
$j$~to individual bodies,
and $k$ to sets of data.
The primed quantities correspond to synthetic data,
integrated (or interpolated) to the times of observations~$t_i$.

We can also add an artificial term:
\begin{equation}
\chi^2_{\rm mass} = \sum_{j=1}^{N_{\rm bod}} \left({2m_j-m_j^{\rm min}-m_j^{\rm max} \over m_j^{\rm max}-m_j^{\rm min}}\right)^{\!\!100}\!\! \label{eq:mass_limit}
\end{equation}
to keep the masses $m_j$ of the components within resonable intervals
(e.g. according to spectroscopic classifications of the components).
The high exponent of the arbitrary function prevents simplex
to drift away from the interval $(m_j^{\rm min}, m_j^{\rm max})$.

\begin{figure}
\centering
\includegraphics[width=10.4cm]{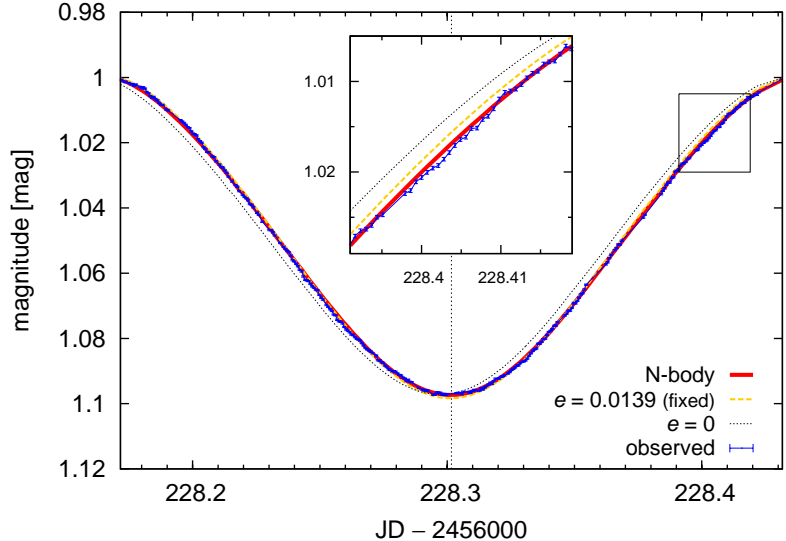}
\caption{Light curves of a detached eclipsing binary and three dynamical models:
(i)~a keplerian (2-body) assuming a fixed circular orbit ($e = 0$, black dotted line),
(ii)~a locally-optimized keplerian with a non-zero fixed eccentricity $e = 0.0139$ (yellow), and
(iii)~a full N-body model with the initial osculating $e_1(t = T_0) = 0$ (red),
but with a general trajectory affected by perturbations among four components.
The last case corresponds to the arrangement of $\xi$~Tauri quadruple system
(as described in Nemravov\'a et al. 2016).
The light curves and minima timings differ
more than the usual uncertainty $\sigma_{\rm lc}$, or
$\sigma_{\rm ttv}$ achievable by space-born observations like that of MOST
(Walker et al. 2003; cf.~blue line with tiny error bars,
quasiperiodic oscillations were removed as explained in Section~\ref{sec:quasi}).
It is thus necessary to use the N-body model for such compact stellar systems,
even on this {\em very short\/} (orbital) time scale.}
\label{fig:lightcurve}
\end{figure}

As usually, observational data have to be in a suitable format
and we provide some example scripts for a conversion or extraction
of data from OIFITS files (Pauls et al. 2005).
Note that one shall {\em not\/} use RV measurements when it is
possible to fit the observed spectra with synthetic ones.
Similarly, no minima timings or durations are needed when we have
complete light curves at disposal (cf.~Figure~\ref{fig:lightcurve});
and no triple product amplitudes $|T_3|$ when the same interferometric
measurements are used as squared visibilities $|V|^2$.
Let us emphasize that it is always better to use directly observable
quantities, not derived!

To find a local or a global minimum, we can use a standard simplex algorithm
or simulated annealing (Nelder \& Mead 1965, Press et al. 1999), with the cooling schedule
${\cal T}^{i+1} = (1-\epsilon_{\rm temp}) {\cal T}^i$,
after given number of iterations at ${\cal T}^i$.
Free parameters of the model (which can be optionally fixed) are:
the masses $m_j$ of the components,
orbital elements $a_j, e_j, i_j, \Omega_j, \omega_j, M_j$ of the respective orbits,
systemic velocity $\gamma$,
distance $d$,
radii $R_j$,
effective temperatures $T_{{\rm eff}\,j}$,
projected rotational velocities $v_{{\rm rot}\,j}$,
and magnitude zero points $m_{0\,k}$.
For $N_{\rm bod}$ bodies, this represents a set of $({\bf 10}N_{\rm bod}+N_{\rm band}-4)$
parameters in total.

Unfortunately, neither of the numerical methods can guarantee
that the global minimum will be found. The multidimensional parameter space is very extended
and there are many local minima, some of them statistically equivalent.
Running simplex many times from different starting points ($10^4$ to $10^5$) can help,
but this problem is clearly both system-dependent and data-dependent.
To obtain uncertainties of the model parameters or full covariance matrix
one can use the bootstrap method (Efron 1979) for example.

\subsection{Dynamical effects}

\begin{figure*}
\centering
\includegraphics[width=14cm]{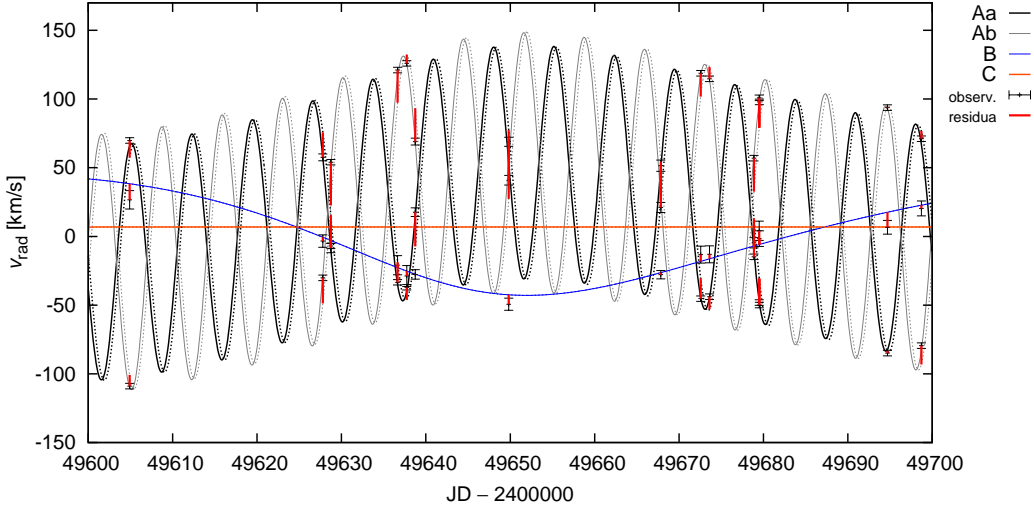}
\caption{An evolution of radial velocities (RVs) of the four components
of $\xi$~Tauri (denoted Aa, Ab, B and C), assuming two different values
of the initial osculating eccentricity~$e_1(t = T_0)$ of the inner orbit:
(i)~zero (thick lines);
(ii)~an increased non-zero $e_1 = 0.01$ (dotted lines).
There is a significant phase shift between them
that can be easily detected, because the respective RV measurements
cover the interval of JD from 2449300 to 2456889.
For even larger $e_1 \simeq 0.1$, the oscillations of RVs forced
by the 3rd body also have larger amplitude, related to the evolution of $e_1(t)$.
For comparison, there are some of the observations plotted (black points
with error bars) and residua wrt.\ the worse non-zero $e_1$ model (red lines).}
\label{fig:forcing}
\end{figure*}

A number of well-known dynamical effects can be modelled with the N-body integrator:
self-consistent precession of $\omega$ and~$\Omega$,
inclination changes and eclipse durations,
eccentricity oscillations (Figure~\ref{fig:forcing}),
Kozai cycles,
variation and evection,
differences between prograde vs retrograde orbits,
close encounters,
hyperbolic trajectories,
mean-motion resonances (Rivera et al. 2005),
secular resonances,
three-body resonances (Nesvorn\'y and Morbidelli 1998), or
chaotic diffusion due to overlapping resonances
are also naturally accounted for in our N-body model.
Even more examples can be found in Fabrycky (2010).

\section{Model testing}

\subsection{A test on synthetic data}

In order to test a basic functionality of our N-body model,
we created a mock system with 40~known parameters; the system
actually closely corresponds to the quadruple star $\xi$~Tau
(see Table~\ref{tab:mock}). Synthetic observational data
were created using the same code with the same coverage and cadence
as the real observations of $\xi$~Tau (Nemravov\'a et al. 2016):
78\,133~spectral measurements (individual data points $I_{\lambda\,i}$),
17\,391~squared visibilities~$|V|^2_i$,
4\,856~complex triple products~$T_{3\,i}$,
2\,974~lightcurve points~$m_{Vki}$,
17~astrometric measurements~$\Delta x_{ij}$, $\Delta y_{ij}$ of the 4th component, and
13~SED points~$m_{Vi}$.
A gaussian noise was applied to all of them, at the levels usual for datasets
we have for $\xi$~Tau; and we assumed there are {\em no\/} systematics,
neither in these synthetic observations, nor in the model
(but cf. Appendices~\ref{sec:heterogenous} to \ref{sec:quasi}).
For the true solution, one would obtain $\chi^2 = 109\,095$,
which is indeed a perfect solution given the number of degrees of freedom
$\nu \equiv N_{\rm data} - M_{\rm free} = 108\,257 - 40 = 108\,217$
and the probability $P(\chi^2|\nu) = 0.970$ that the $\chi^2$ value is that
large by chance.

\begin{figure*}
\centering
\includegraphics[width=8.5cm]{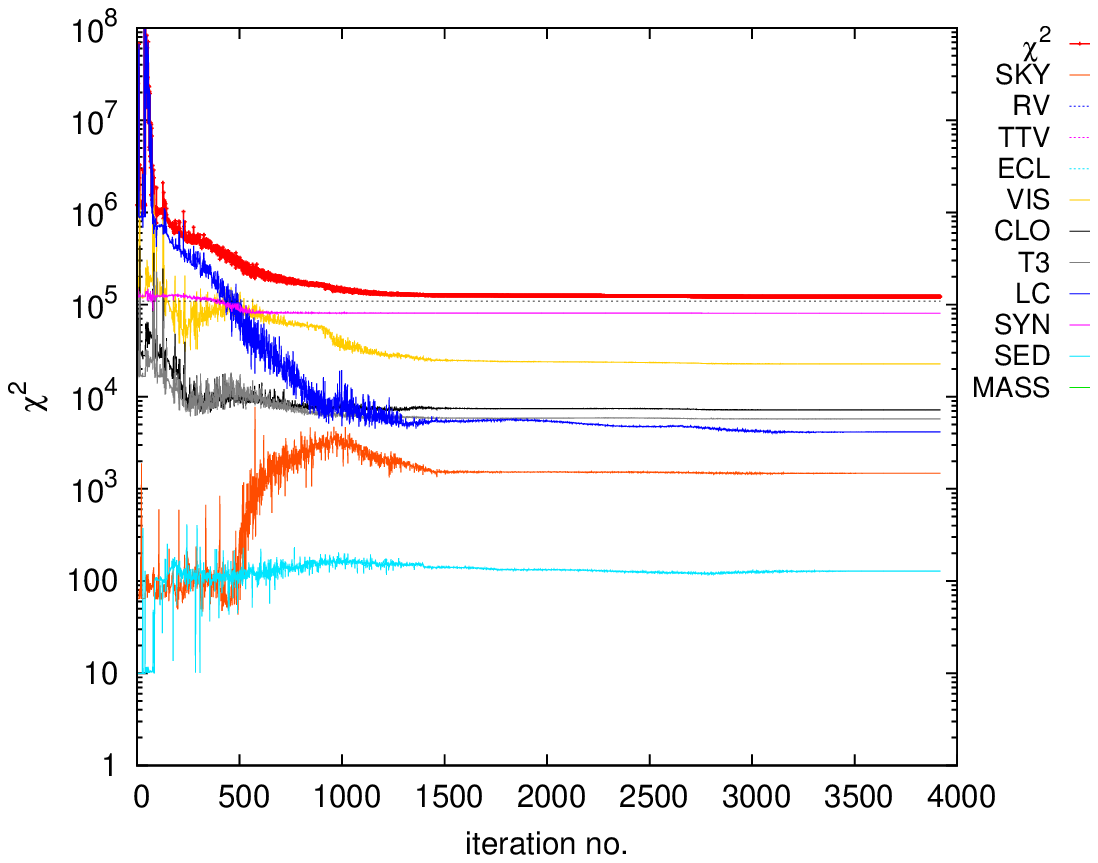}
\includegraphics[width=8.5cm]{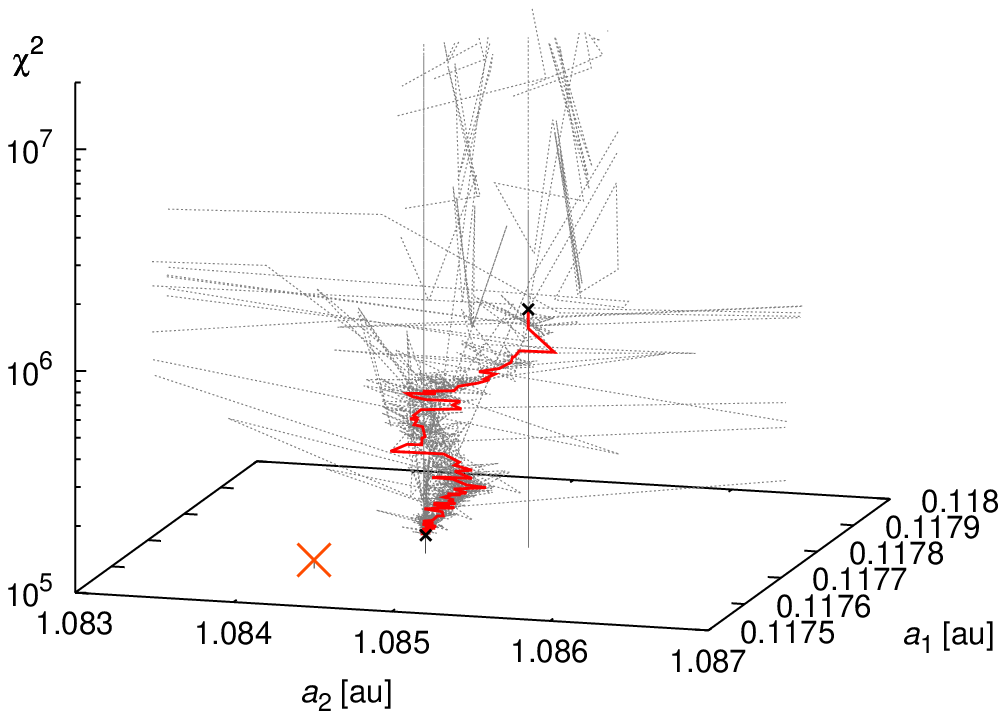}
\caption{Left: A convergence of the simplex to a local minimum
vs the number of iterations
for the mock system from Table~\ref{tab:mock}.
Individual contributions to the total $\chi^2$
corresponding to Eq.~(\ref{eq:chi2}) are shown.
Initially, all of the 40 parameters were shifted by $0.1\,\%$
and the $\chi^2$ value suddenly increased up to $1.19\times10^7$
because the model is very sensitive to some of them.
The final value after ${\simeq}\,10^3$ iterations is $\chi^2 = 120\,625$.
This is quite close to the true solution, but still some restarts
of the simplex (or simulated annealing) would be needed to obtain
$\chi^2$ as low as $109\,095$, i.e. the value of the true solution
(albeit with noisy synthetic data). Moreover, iterations with increased
weigths $w_{\rm sky}$ and $w_{\rm sed}$ would be also needed.
Right: The same convergence of $\chi^2$ with respect to semimajor axes $a_1$, $a_2$
(i.e. 2 out of 40 free parameters).
One may see the initial (offset) and final positions (black crosses),
sucessful steps (red solid lines),
unsucessful trials (gray dotted),
and the true solution (orange cross).
While the simplex approaches the true solution,
it is often stuck half-way in a local minimum.}
\label{fig:chi2_iter}
\end{figure*}

\begin{figure}
\centering
\includegraphics[width=8.5cm]{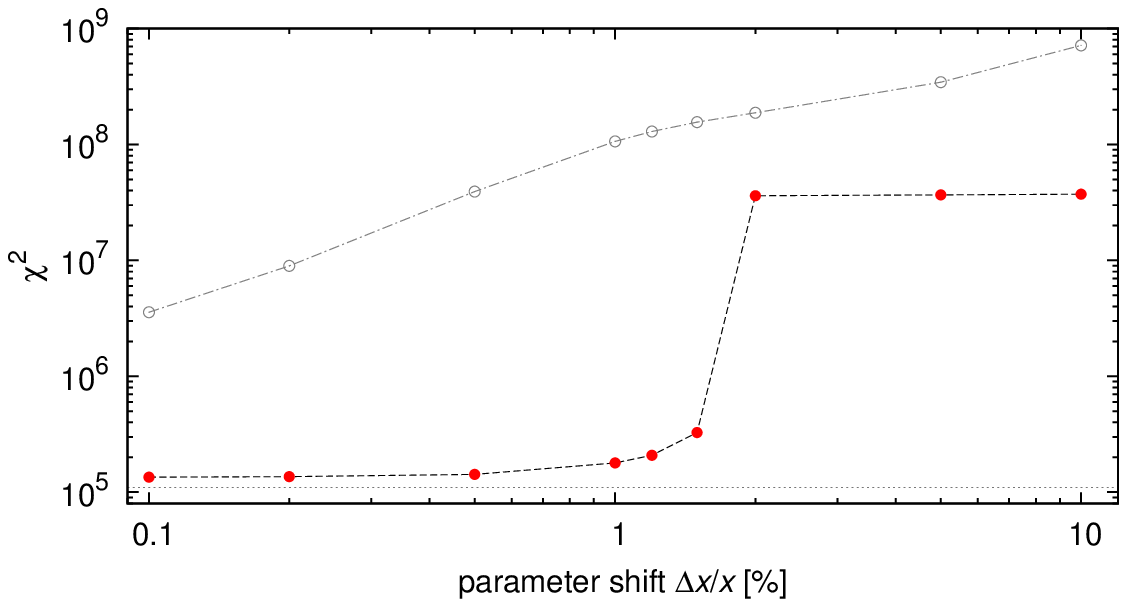}
\caption{Final $\chi^2$ values (filled circles) after ${\simeq}\,10^3$ simplex iterations
vs relative shifts of the initial parameters (expressed in percentages).
For comparison, there are also initial $\chi^2$'s (open circles) we started simplex at.
The true solution reference value is $\chi^2 = 109\,095$ (dotted line).
For the purpose of this test, we used osculating periods $P_i$
as fixed (and unshifted) parameters, instead of free semimajor axes $a_i$,
as they are usually well constrained by period analyses.
We also kept fixed $T_{{\rm eff}i}$, $v_{{\rm rot}\,i}$
and used simple bandpasses and Planck approximation to speed-up computations.}
\label{fig:chi2_shift}
\end{figure}

We then performed several simplex optimisations, starting from a number
of neighbouring points, farther and farther away from the true solution.
The convergence of the simplex to a local minimum is clear (Figure~\ref{fig:chi2_iter})
and we recovered the original parameters (low $\chi^2 \simeq \nu$)
with some uncertainties, as expected; unless the initial guess was too far away,
say more than a few percent of the critical parameters
(see Figure~\ref{fig:chi2_shift}). More extended surveys
with many initial starting points and/or simulated annealing would be
needed in these unfortunate cases. The differences between the final
and true solutions roughly correspond to the uncertainties we would
obtain from bootstrap testing.

Often some preliminary knowledge based on observation-specific models
is available (e.g. prominent periods, previously published parameters).
The N-body model is especially suitable for such `semi-final' convergence
--- with all parameters free --- which should limit the systematics arising
from a usage of limited (keplerian) models.

Regarding the fractional or missing data, there are several rather
trivial facts, e.g. if we miss RVs, it is impossible to resolve
low-$e$ orbits from high-$e$ with $\omega = 0^\circ$ or $180^\circ$.
If there are no eclipses and no interferometric measurements available, one cannot
precisely constrain the inclination~$i$. Without closure phase measurements,
there is practically no sensitivity to asymmetries, etc.
Of course, they are interesting when dealing with real observational datasets
(see also Appendix \ref{sec:mirror}).

Let us point out that this kind of testing has somewhat limited capabilities.
First, if we create the mock data with the same model, then this test
is essentially the test of the numerical methods (simplex and simulated
annealing). Because these methods are indeed classical (Nelder \& Mead 1969),
their limitations are already very well known.

Second, to create the mock data independently, one would need a completely
independent model with exactly the same capabilities. However, this is
rather a test of systematic differences between the models and not
of the method itself. We consider this approach to be more useful,
but it is rather difficult to obtain the second model.
Of course, any keplerian models (e.g. Phoebe 1.0) are useless.
Any model which does not produce all the observables
(astrometry, RVs, TTVs, $\epsilon_{\rm ecl}$, $|V|^2$, $T_3$, $m_V$, or $I_\lambda$)
is impractical too, because we need as much orthogonal constraints as possible.
Analytical photometric models (e.g. Carter et al. 2008) are too simplified
for stellar binaries. And so on. It may be possible to use Phoebe 2.0
(Pr\v sa et al. 2016) for such comparison in the future.

Last but not least, real observational data of real systems have its own cadence,
coverage, calibrations, uncertainties, and systematics. Even though
one can play with artificial data, these tests cannot be used
straightforwardly, because we know "nothing" a-priori about
the given data we obtain from observers.
Consequently, one will have to perform suitable tests again and again
for every next system.

\begin{table*}
\caption{The mock system parameters which were used to generate
synthetic data for testing. The notation is the same as in Table~\ref{tab:quantities}.
Values rounded to typical uncertainties of the parameters are presented
in this table. The osculating elements correspond to the epoch $T_0 = 2456224.724705$.}
\label{tab:mock}
\centering
\begin{tabular}{lllllllll}
\hline
\hline
Par. & Value & & & & & & & Unit \\
\hline
$m_1$ & $2.238483$ &
$m_2$ & $2.009645$ &
$m_3$ & $3.7472  $ &
$m_4$ & $0.92    $ &
$M_\odot$ \\
$a_1$ &  $0.1176356$ &
$a_2$ &  $1.08420  $ &
$a_3$ &  $28.39    $ &
& & au \\
$e_1$ &  $0.0000$ &
$e_2$ &  $0.2167$ &
$e_3$ &  $0.568 $ &
\\
$i_1$ &  $ 87.6$ &
$i_2$ &  $ 86.3$ &
$i_3$ &  $-18.2$ &
& & deg \\
$\Omega_1$ & $ 329.1$ &
$\Omega_2$ & $ 328.6$ &
$\Omega_3$ & $ 114.7$ &
& & deg \\
$\omega_1$ & $ 275.65$ &
$\omega_2$ & $   0.00$ &
$\omega_3$ & $   1.0 $ &
& & deg \\
$M_1$ & $174.44$ &
$M_2$ & $ 88.04$ &
$M_3$ & $ 32.7 $ &
& & deg \\
$\gamma$ & $8.82$ &
& & & & & & ${\rm km}\,{\rm s}^{-1}$ \\
$d$ & $67.9$ &
& & & & & & pc \\
$T_{\rm eff1}$ & $10727$ &
$T_{\rm eff2}$ & $10275$ &
$T_{\rm eff3}$ & $13120$ &
$T_{\rm eff4}$ & $6526$  &
K \\
$R_1$ & $1.586$ &
$R_2$ & $1.642$ &
$R_3$ & $2.727$ &
$R_4$ & $0.877$ &
$R_\odot$ \\
$v_{\rm rot1}$ & $ 16.2$ &
$v_{\rm rot2}$ & $ 12.5$ &
$v_{\rm rot3}$ & $234.9$ &
$v_{\rm rot4}$ & $ 80.1$ &
${\rm km}\,{\rm s}^{-1}$ \\
$m_{01}$ & $1.000$ &
$m_{02}$ & $3.335$ &
$m_{03}$ & $3.646$ &
$m_{04}$ & $3.730$ &
mag \\
\hline
\end{tabular}
\vskip1cm
\end{table*}

\subsection{A comparison with other models}

As already mentioned in the Introduction, a comparison with
several observation-specific models was already done
in our previous paper Nemravov\'a et al. (2016).
In particular, our N-body model produces results which
are compatible {\em within\/} respective approximations
(e.g. on short time scales when the keplerian model
can be regarded as a useful approximation) with the following
published works:
photometric model of Phoebe 1.0 (Pr\v sa \& Zwitter 2006);
astrometric and speckle-interferometry model of Zasche \& Wolf (2007);
RV disentangling by Korel (Hadrava 1997);
LitPro model for visibilities $|V|^2$ (Tallon-Bosc et al. 2008);
spectro-interferometric $|V|^2$ and $T_3$ (Nemravov\'a et al. 2016);
synthetic spectra fitting by Pyterpol (dtto).


\section{Conclusions and future work}\label{sec:conclusions}

Today, N-body models seem to be an absolutely necessary tool
for a careful inspection of observational data.
It is important to take care that discrepancies between keplerian
and full N-body dynamics no longer spoil derived stellar parameters.
After a removal of (some) systematic errors (sometimes) present in observations or reductions,
it enables us to reveal even tiny N-body perturbations
and construct robust models of compact stellar systems
(e.g.~those from Table~\ref{tab:systems}).

Regarding future developments of (our or other) N-body models,
it seems worthwhile to also account for:
calibration factors of individual interferometric telescopes,
gravity darkening in the visibility calculation of rotating stars (as in Aufdenberg et al. 2006),
especially when measuring on longest baselines, 
and eventually one may think of an upgrade to the WD 2015, or Phoebe 2.0
(already used in Pablo et al. 2015).

Yet another work is needed to compute trajectories even more accurately,
with physics going beyond point-like masses, equilibrium tides or oblateness, namely:
higher gravitational moments ($J_4$) due to non-sphericity of stellar components,
tidal dissipation and cross tides (e.g.~Mignard 1979),
corresponding long-term evolution of orbits,
spin evolution (Eggleton \& Kisileva-Eggleton 2001),
spin--orbital resonances,
or radiation of gravitational waves in extreme cases.

The situation in stellar interiors also matters. The dissipation
occurs either due to viscosity in outer convective zones,
or due to inertial oscillations in radiative zones,
which are excited on eccentric orbits by dynamic tides
and subsequently radiatively damped (Zahn 2008).
In triple systems, the excitations may actually arise from
a~binary subsystem, and corresponding light oscillations
then have an~half of its period (Derekas et al. 2011, Fuller et al. 2013).
Another difficulty stems from certain coupling of envelopes
and cores (Papaloizou \& Ivanov 2010).
Inevitably, a fully self-consistent model should account
for a back-reaction: the strongest tidal heating may inflate
whole objects (Mardling 2007).

\begin{table}
\caption{Suggested examples of compact stellar systems for which the N-body model
could be useful (or inevitable). This `catalogue' obviously cannot be considered
comprehensive.}
\label{tab:systems}
\centering
\begin{tabular}{ll}
\hline
\hline
Designation & Reference \\
\hline
$\lambda$~Tau             & Fekel \& Tomkin (1982) \\
$\xi$~Tau                 & Nemravov\'a et al. (2016) \\
VW~LMi                    & Pribulla et al. (2008) \\
V994 Her = HD 170314      & Zasche \& Uhl\'a\v r (2016) \\
V907 Sco = HD 163302      & Lacy et al. (1999) \\
HD 91962                  & Tokovinin et al. (2015) \\
HD 109648                 & Jha et al. (2000) \\
HD 144548                 & Alonso et al. (2015) \\
HD 181068 = KIC 5952403   & Fuller et al. (2013) \\
KIC 05255552              & Borkovits et al. (2016) \\[-4pt]		
KIC 05771589              & \\[-4pt] 					
KIC 06964043              & \\[-4pt]					
KIC 07289157              & \\[-4pt]					
KIC 07668648              & \\[-4pt] 					
KIC 07955301              & \\[-4pt]					
KIC 09714358              & \\						
\hline
\end{tabular}
\end{table}

\acknowledgments
The work of MB was supported by the grants no. P209-15-02112S and P209-13-01308S
of the Czech Science Foundation.
I thank Jana Nemravov\'a and David Vokrouhlick\'y for valuable discussions
on the subject and a fruitful collaboration on the $\xi$~Tauri paper.
I also have to thank the referee Hagai Perets for a constructive criticism
which determined the final structure of the paper.


\appendix

\section{Possible problems due to systematics}\label{sec:problems}

We have to admit that any modelling (compact stellar systems included)
can be spoiled, either when there are systematic deficiencies of the model,
e.g. keplerian vs N-body, or serious systematic errors in observational data,
especially when we use very heterogeneous datasets.
In the following, we thus discuss several `dangerous' cases.

\subsection{Discretization errors}

Of course, any numerical computation suffers from discretization
errors and interpolation errors, even though we tried to decrease
the latter as much as possible (cf.~Section~\ref{sec:model}).
This is probably the most important disadvantage compared to analytical
computations. A general rule is a convergence of results
(and corresponding $\chi^2$ values) for $\Delta t \to 0$.

However, let us add a warning that rarely a decrease of time step,
e.g. by a factor of~2, may lead to unexpected results.
For example, when eclipses are almost disappearing,
the trajectory with $\Delta t/2$ is more curved and
may thus miss the last eclipse, which suddenly {\em increases} $\chi^2_{\rm ttv}$
because the next eclipse is now one orbital period~$P$ far away.
The solution is to converge the model once again, with~$\Delta t/2$.

Let's not forget, there is yet another discretization related
to the WD code, or the surfaces of the eclipsing binary.
For low numbers $N_{\rm wd}$, one can see numerical artefacts
on the light curve, as rectangular surface facets appear
from behind the limb, or disappear. Again, it is worth to check
larger~$N_{\rm wd}$.

\subsection{Mirror solutions}\label{sec:mirror}

Quite often, we can expect one or more ($m$) mirror solutions
(and $2^m$ combinations of them). A typical situation is we have
no RVs for faint components (so that both inclinations $i$ and $i' = -i$
are admissible), or no unambiguous astrometry or closure-phase measurements
(so that $\Omega$ and $\Omega' = 180^\circ-\Omega$ are both admissible).
Consequently, one may save some time when surveying the parameter space.

However, with the N-body model at hand it is worth to check not only
the total $\chi^2$ but also {\em individual contributions\/} to~$\chi^2$
for all the mirror models! Especially~$\chi^2_{\rm ttv}$
is very sensitive to the mutual perturbations, and we may be able
to resolve some of the ambiguities mentioned above.

Of course, the statistics must not be corrupted by systematics or strongly
underestimated uncertainties in other observational datasets.
If this is the unfortunate case, one may try to use weights~$w$
of individual $\chi^2$'s, but this should be used as ``a method of last resort''.
The reason is that it is too easy to hide {\em all\/} systematics this way,
even though it is better to get rid of them (see below).

\subsection{Heterogeneous datasets of RVs}\label{sec:heterogenous}

Radial-velocity measurements might be affected by zero point offsets,
which then lead to different systemic velocities $\gamma$
for different observatories. This can be a bit misleading,
because it is not possible to {\em a priori\/} distinguish systematic differences
in dispersion relations from real perturbations,
when the observations were acquired at epochs distant in time.

A well-known viable approach is to use an independent calibration
by narrow interstellar lines (DIBs; Chini et al. 2012), if they are present
and resolved in the given spectral range. Another possibility
are atmospheric lines for which the relative RVs can be computed easily.
If this is impossible, one should use the N-body model with a great
caution, because simply increasing $\sigma_{\rm rv} \simeq \Delta\gamma$,
to get $\chi^2_{\rm rv} \simeq N_{\rm rv}$ is a wrong idea.
The RV measurements in question will still `push' the model elsewhere
and there will be systematic departures with respect to other (more or less orthogonal) observational data.

It may be a too much freedom, but if the dispersion relations can be
considered stable from night to night, some calibration factors $f_{{\rm rv}\,k}$ of RVs
--- assigned to individual observatories or datasets --- might be actually a better solution.
In any case, such factors have to be always treaded as additional
free parameters of the N-body model.

\subsection{RVs from disentangling}

Sometimes, RVs are derived in the Fourier domain
by means of disentangling (e.g. by Korel; Hadrava 1995),
with an advantage to obtain disentangled spectra of individual components.
There is a `hidden' caveat, though, because one can expect
a strong correlation of RVs and the fixed keplerian orbital elements
used during the disentangling procedure. This represents a problem,
because we do vary initial osculating orbital elements
in the N-body model and they most likely will contradict
the previous elements.

Note the disentangled spectra should {\em not\/} be re-used as templates,
because they contain slight systematic asymmetries or wavy continua.
If we try to match the observed spectra with such templates again,
we would obtain artificially small uncertainties $\sigma_{\rm rv}$
(and extremely large $\chi^2_{\rm rv}$).
A~solution is to use synthetic spectra similar to the disentangled ones,
but with no direct relation to Korel,
as an intermediate step to derive new RVs.

\subsection{RVs from synthetic spectra}

Alternatively, RVs of the individual components can be derived directly
in the time domain by fitting a luminosity-weighted sum of suitable
synthetic spectra (e.g. by Pyterpol; Nemravov\'a et al. 2016).
Instead of fitting the observed spectra individually (one-by-one), 
it is advisable to assume that most of the free parameters
(projected $v_{{\rm rot}\,j}$, $T_{{\rm eff}\,j}$, gravity~$\log g_j$,
and metallicity~$Z_j$ of the stellar components)
are the same for all spectra, with the exception of RVs which are surely time dependent.
Luckily, these RVs are {\em not\/} strongly correlated with the orbital elements,
so they seem suitable as an input for the N-body model.

\begin{figure}
\centering
\includegraphics[width=12cm]{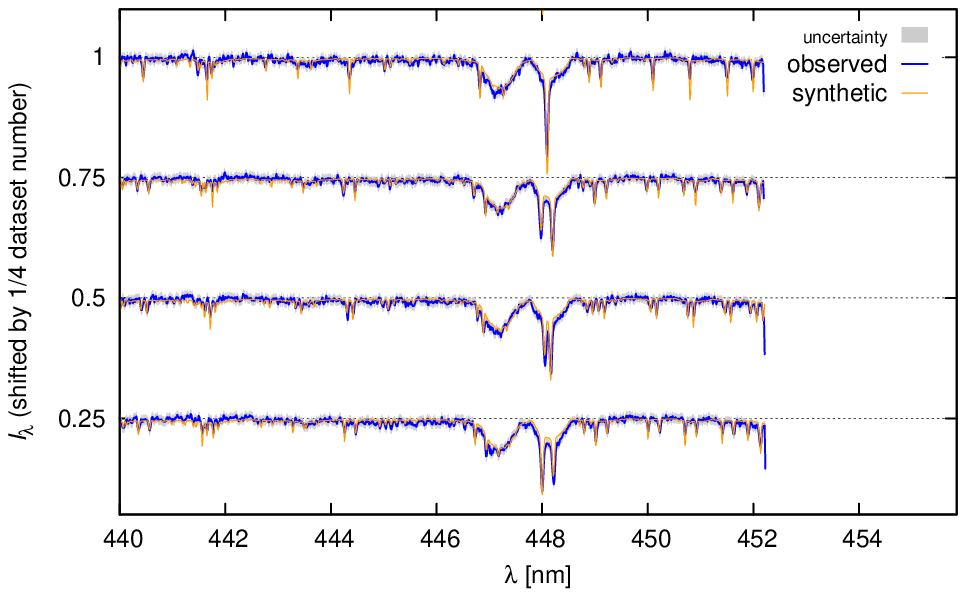}
\caption{A small subset of the observed spectra of $\xi$~Tauri (blue),
fitted by a triplet of synthetic spectra (orange) for components Aa, Ab (sharp-lined), and B (broad-lined).
The Doppler shifts were set according to the N-body model,
consequently there is no problem with the blending of lines in the top spectrum.
The respective parameters of the components were assumed as follows:
the effective temperature $T_{\rm eff} = 10700,\, 10480,\, 14190\,{\rm K}$;
surface gravity $\log g = 4.08,\, 4.01,\, 4.527$;
projected rotational velocity $v_{\rm rot} = 12.6,\, 14.3,\, 229.2\,{\rm km}\,{\rm s}^{-1}$; and
the metallicity was solar.
The relative luminosities were $L = 0.203,\, 0.134,\, 0.644\,L_\odot$,
while the component C was considered too faint.
The synthetic spectra were prepared with Pyterpol (Nemravov\'a et al. 2016).}
\label{fig:synthetic}
\end{figure}

On the other hand, this method can have problems on its own when RVs are small (at conjunctions)
and $v_{\rm rot}$ large, so that the lines are totally blended.
As a provisional solution, one may try to discard the lowest RVs which cause the problems,
or do {\em not\/} use RVs at all and rather fit synthetic spectra directly
with the N-body model ($\chi^2_{\rm syn}$), which is definitely a better approach,
because RVs will be correctly tied to each other (see Figure~\ref{fig:synthetic}).

\subsection{Rectification procedure}

Inevitably, RVs might have been systematically affected already during a basic reduction,
namely a rectification (normalisation) of spectra. If the rectification procedure
is automated by fitting a low-degree polynomial to continua, it is worth to try
a different maximum degree of the polynomial and run the above synthetic spectra
optimisation once again.

\subsection{Visibility calibration}

Contrary to closure phase $\arg T_3$ measurements, the squared visibility~$|V|^2$
has to be calibrated by close-in-time observations of comparison stars
with known angular diameters or unresolved (point-like) sources.
Sometimes even the calibrated measurements exhibit unrealistically quick
changes of $|V|^2$ or sudden decreases of $|V|^2$, possibly caused by
unfavorable weather conditions, or seeing comparable to the slit width,
affecting a light contribution from barely-resolved components,
or other obscure instrumental defects.

In the end, dropping of these suspicious observational data may be the
only way to prevent the systematics to unrealistically shift the model.
Using a low weight $w_{\rm vis} = 0.1$ is not a satisfactory option.
To this point, we always retain a dataset identification for each single measurement
which enables us to quickly perform a bootstrap testing. 

\subsection{Quasiperiodic oscillations}\label{sec:quasi}

A removal of quasiperiodic light oscillations which are sometimes present
(or always for high-precision measurements)
outside eclipses is very important, because they may
otherwise systematically offset the minima timings themselves.
One wave of the oscillations behaves like a `ramp', which skews
the light curve at around the minimum.

The observed light curve should be thus {\em locally\/} fitted by a suitable function
(e.g. harmonic with a variable period and amplitude) and then
subtracted from the data. If the (synthetic) light curve out of eclipses
is flat beyond doubt, it seems better to drop these segments of the (observed)
light curve completely, because they would increase $\chi^2_{\rm lc}$ but there is no useful information
as we have no physical model for these oscillations (as of yet).

\subsection{Osculating vs fixed elements}

Some care is also needed when comparing results of (old) keplerian and (new) N-body models.
They actually {\em can\/} differ by more than a few $\sigma$,
because the former orbital elements are fixed,
while the latter are only osculating initial conditions at $t = T_0$.
Generally, all elements are time-dependent quantities, $a_1(t), e_1(t), i_1(t),$ etc.,
whereas their oscillations are often {\em larger\/} than uncertainties
of the initial osculating elements.
In fact, one can perform some averaging over the observational time span
to facilitate the comparison.
Nevertheless, the N-body model is more complete, and it should be probably preferred.

\subsection{Stability, aliasing, mean and proper elements}

It is also possible to run the N-body integrator separately,
regardless of an observational time span, and study a long-term evolution
and stability of stellar systems. We may wish to prefer those
orbital solutions which are indeed stable.
One of the difficulties is that the output of osculating elements is either
prohibitively long or an {\em aliasing\/} occurs when the output
time step $\Delta t_{\rm out}$ is larger than an half of the shortest
orbital period, $P_1/2$.

In a modified version of the BS integrator (swift\_bs\_fp),
we can use an on-line digital filtering of non-singular osculating elements
$h_j, k_j, p_j, q_j$ to overcome these problems:
first a multi-level convolution based on the Kaiser windows (Quinn et al.\ 1991)
to obtain {\em mean\/} elements, and
second a frequency-modified Fourier transform (\v Sidlichovk\'y \& Nesvorn\'y 1997)
to extract {\em proper\/} elements. For~$N$ mutually interacting bodies,
one can expect $2N$~eigen-frequencies of the system, which are usually denoted
$g_j$ and $s_j$. The corresponding amplitudes $e_{{\rm p}j}$, $\sin {1\over 2}I_{{\rm p}j}$
can be considered approximate integrals of motion which only evolve
on time scales longer than secular.

\vskip\baselineskip

To conclude in a pessimistic way, the above list of possible problems
and systematics cannot be treated as complete, unfortunately.


\section{Technical notes}

\subsection{Different hierarchy}

By default, we assume a hierarchy like ((1+2)+3)+4,
for which Jacobian orbital elements seem to be a suitable description.
For a substantially different hierarchy,
say two pairs like (1+2) and (3+4),
where we would prefer a different definition of elements,
only a very small part of the code has to be rewritten,
namely in the geometry.f subroutine, where the elements
are converted to barycentric Cartesian coordinates.
Alternatively, one may wish to use 1-centric Cartesian
coordinates as actual parameters, because sometimes precise observational
data may constrain (`fix') some of them ($x_{{\rm h}2}$, $y_{{\rm h}2}$, etc.),
thus decreasing the dimensionality of the parameter space.

\subsection{Jacobian orbital elements}

Unlike the usual stellar-astronomy convention, where the brightest component
is always at the origin of the reference frame, in our N-body model we usually
select the most compact eclipsing pair as bodies 1 and 2, or the most massive
component as~1. The reason is that orbital elements in hierarchical systems
are usually computed in Jacobian coordinates, where the centre of mass 1+2
is the reference point for the coordinates and velocities of the 3rd body;
the 1+2+3 centre of mass is a suitable reference for the 4th body, and so on.
The corresponding Jacobian elements then have a nice interpretation.
Because of the above definition, it may be necessary to adjust
to-be-fitted astrometric measurements by $180^\circ$ in the position angle
--- not due to an ambiguity, but simply because the reference body is different
in our case. Similarly, a~value of~$\Omega$ from literature
may actually differ by $180^\circ$.

\end{document}